\documentclass[]{aa}  

\usepackage{graphicx}
\usepackage{txfonts}
\usepackage{graphicx}
\usepackage{txfonts}
\usepackage{graphicx}
\usepackage{lscape}
\usepackage{longtable}
\usepackage{natbib}
\usepackage{color}
\usepackage{array} 
\usepackage{array} 
\usepackage{tikz,array}
\usetikzlibrary{calc}
\usepackage{txfonts}
\usepackage{color}
\usepackage{multirow}
\usepackage{here}

\usepackage{txfonts}
\usepackage{amsmath,amstext}
\usepackage{graphicx}
\usepackage{epstopdf}
\usepackage{float}
\usepackage{array}
\usepackage{mathtools}
\usepackage{booktabs}
\usepackage{subfigure}
\usepackage{url}
\usepackage{helvet}
\usepackage{tabularx}
\usepackage{multirow}
\usepackage{natbib}
\usepackage[flushleft]{threeparttable}
\usepackage{lscape}
\usepackage{pdflscape}
\usepackage{longtable}
\usepackage{wasysym}
\usepackage{float}
\usepackage{relsize}
\usepackage{color}
\usepackage{breqn}
\usepackage{bm}
\usepackage{enumitem}

\usepackage[colorlinks, citecolor=blue, linkcolor=blue]{hyperref}
\hypersetup{colorlinks,breaklinks, linkcolor=blue,urlcolor=magenta, anchorcolor=blue,citecolor=blue}

\def\ms{\hbox{m s$^{-1}$}}         
\def\gcm3{\hbox{g cm$^{-3}$}}       
\def\Msun{\hbox{$\mathrm{M}_{\astrosun}$}}             
\def\Rsun{\hbox{$\mathrm{R}_{\astrosun}$}}

\def\Mearth{\hbox{$\mathrm{M}_{\oplus}$}}
\def\Rearth{\hbox{$\mathrm{R}_{\oplus}$}}

\bibpunct{(}{)}{;}{a}{}{,} 

\newcommand{\be}{\begin{equation}}
\newcommand{\ee}{\end{equation}}

\usepackage{romannum}

\begin{document}

   \title{The KOBE experiment: \\K-dwarfs orbited by habitable exoplanets
   \thanks{Based on observations collected at Centro Astron\'omico Hispano en Andaluc\'ia (CAHA) at Calar Alto, operated jointly by Instituto de Astrof\'isica de Andaluc\'ia (CSIC) and Junta de Andaluc\'ia}}

   \subtitle{Project goals, target selection, and stellar characterization}

   \author{
J.~Lillo-Box\inst{\ref{cab}}, 
N.~C.~Santos\inst{\ref{ia-porto},\ref{uporto}},
A.~Santerne\inst{\ref{marseille}},
A.~M.~Silva\inst{\ref{ia-porto},\ref{uporto}},
D.~Barrado\inst{\ref{cab}},
J.~Faria\inst{\ref{ia-porto},\ref{uporto}},
A.~Castro-Gonz\'alez\inst{\ref{cab}},
O.~Balsalobre-Ruza\inst{\ref{cab}},
M.~Morales-Calder\'on\inst{\ref{cab}},
A.~Saavedra\inst{\ref{cab}},
E.~Marfil\inst{\ref{cab}},
S.~G.~Sousa\inst{\ref{ia-porto}},
V.~Adibekyan\inst{\ref{ia-porto},\ref{uporto}},   
A.~Berihuete\inst{\ref{ucadiz}},
S.~C.~C.~Barros\inst{\ref{ia-porto},\ref{uporto}},
E.~Delgado-Mena\inst{\ref{ia-porto},\ref{uporto}},
N.~Hu\'elamo\inst{\ref{cab}},
M.~Deleuil\inst{\ref{marseille}},
O.~D.~S.~Demangeon\inst{\ref{ia-porto},\ref{uporto}}, 
P.~Figueira\inst{\ref{geneva}},
S.~Grouffal\inst{\ref{marseille}}
J.~Aceituno\inst{\ref{caha}},
M.~Azzaro\inst{\ref{caha}},
G.~Bergond\inst{\ref{caha}}, 
A.~Fern\'andez-Mart\'in\inst{\ref{caha}}, 
D.~Galad\'i\inst{\ref{caha}},
E.~Gallego\inst{\ref{caha}}, 
A.~Gardini\inst{\ref{caha}},
S.~G\'ongora\inst{\ref{caha}},
A.~Guijarro\inst{\ref{caha}}, 
I.~Hermelo\inst{\ref{caha}}, 
P.~Mart\'in\inst{\ref{caha}},
P.~M\'inguez\inst{\ref{caha}},
L.M.~Montoya\inst{\ref{caha}}, 
S.~Pedraz\inst{\ref{caha}}, 
J.~I.~Vico~Linares\inst{\ref{caha}}, 
}

\institute{
Centro de Astrobiolog\'ia (CAB, CSIC-INTA), Depto. de Astrof\'isica, ESAC campus, 28692, Villanueva de la Ca\~nada (Madrid), Spain \email{Jorge.Lillo@cab.inta-csic.es  }
\label{cab}
\and
Instituto de Astrof\'isica e Ci\^encias do Espa\c{c}o, Universidade do Porto, CAUP, Rua das Estrelas, 4150-762 Porto, Portugal
\label{ia-porto}
\and
Departamento de F\'isica e Astronomia, Faculdade de Ci\^encias, Universidade do Porto, Rua do Campo Alegre, 4169-007 Porto, Portugal
\label{uporto}
\and
Aix Marseille Univ, CNRS, CNES, LAM, Marseille, France
\label{marseille}
\and
Depto. Estad\'istica e Investigaci\'on Operativa, Universidad de C\'adiz, Avda. Rep\'ublica Saharaui s/n, 11510, Puerto Real, C\'adiz
\label{ucadiz}
\and
Observatoire astronomique de l'Universit{\'e} de Gen{\`e}ve, Chemin Pegasi 51, 1290 Versoix, Switzerland
\label{geneva}
\and
Centro Astron\'omico Hispano en Andaluc\'\i a, Observatorio de Calar Alto, Sierra de los Filabres, 04550 G\'ergal, Almer\'\i a, Spain
\label{caha}
}

   \date{in prep.}

 
\abstract{The detection of habitable worlds is one of humanity's greatest endeavors. Thus far, astrobiological studies have shown that one of the most critical components for allowing life to develop is liquid water. Its chemical properties and its capacity to dissolve and, hence, transport other substances makes this constituent a key piece in this regard. As a consequence, looking for life as we know it is directly related to the search for liquid water. For a remote detection of life in distant planetary systems, this essentially means looking for planets in the so-called habitable zone. In this sense, K-dwarf stars are the perfect hosts to search for planets in this range of distances. Contrary to G-dwarfs, the habitable zone is closer, thus making planet detection easier using transit or radial velocity techniques. Contrary to M-dwarfs, stellar activity is on a much smaller scale, hence, it has a smaller impact in terms of both the detectability and the true habitability of the planet. Also, K-dwarfs are the quietest in terms of oscillations, and granulation noise. In spite of this, there is a dearth of planets in the habitable zone of K-dwarfs due to a lack of observing programs devoted to this parameter space. In response to a call for legacy programs of the Calar Alto observatory, we have initiated the first dedicated and systematic search for habitable planets around these stars: K-dwarfs Orbited By habitable Exoplanets (KOBE). This survey is monitoring the radial velocity of 50 carefully pre-selected K-dwarfs with the CARMENES instrument over five semesters, with an average of 90 data points per target. Based on planet occurrence rates convolved with our detectability limits, we expect to find $1.68\pm 0.25$ planets per star in the KOBE sample. Furthermore,  in half of the sample, we expect to find one of those planets within the habitable zone. Here, we describe the motivations, goals, and target selection for the project as well as the preliminary stellar characterization.}

   \keywords{Planets and Satellites: detection, fundamental parameters -- Techniques: radial velocities }

        \titlerunning{The KOBE experiment}
        \authorrunning{Lillo-Box et al.}

   \maketitle
%

\section{Introduction}

In our hunt for Earth analogs, extrasolar planets have been searched extensively in the past few decades (e.g., \citealt{borucki10}), with the focus on detecting as many of them as possible, regardless of their properties. These studies have largely been biased and limited by the instrumental precision. Thanks to all these prior efforts (including ground- and space-based observations), more than 5000 planets are currently known. Their properties are extremely diverse, demonstrating characteristics that are quite distant from those expected from the components of the Solar System and, initially, even from those envisioned by theories of planet formation \citep[see, e.g.,][]{udry07,mayor14}. 

One of the main goals of exoplanet exploration is the search for new planets with the capabilities to develop and sustain life. In this regard, water plays a key role. Its physical and chemical properties make this constituent
a key piece in the development of life as we know it, namely: the ability to form hydrogen bonds to create a sticky liquid that aggregates and stays together, property of expansion when frozen, broad temperature range over which it remains liquid, high heat capacity and high boiling temperature, surface tension, polarity, and capacity to dissolve and, hence, transport other substances in hydrologic cycles. Therefore, in practice, the search for life in the Universe translates into a search for rocky worlds capable of retaining liquid water \citep{kasting93,kopparapu13}.

In the Solar System, this "follow the water" premise undertaken by the National Aeronautics and Space Agency (NASA) under the Mars Exploration Program\footnote{See \url{https://www.nasa.gov/pdf/168049main_Follow_the_Water.pdf}} has driven the focus of the exploration of Mars and the moons of the gas giants Jupiter and Saturn in past decades and in our day. In the exoplanetary context, and given the current capabilities, this premise corresponds to the search for rocky planets in the so-called habitable zone, the range of distances from the star in which water could be in a liquid state on the surface of a rocky world \citep{kopparapu13}.

The search for habitable-zone worlds has first focused on stellar properties similar to those of the Sun (e.g., \citealt{borucki10}, \citealt{pepe11}, \citealt{pepe20,hojjatpanah19}, \citealt{villanueva21}). In recent years, M-dwarfs have been intensively targeted (e.g., \citealt{bonfils05}; \citealt{quirrenbach14}, \citealt{lillo-box20}, , \citealt{demangeon21}) because their habitable zone is closer to the star and the stars themselves are less massive, thus making the detection of habitable Earth analogs (and any planet in general) easier using transits or radial velocity techniques. Both regimes, while very relevant, suffer from substantial difficulties, from both the instrumentation point of view and from the physical side. 

M-dwarfs are very active \citep{jeffers18}, which means the radial velocity analysis faces great challenges in detecting planet-like signals at the meter-per-second level. Also, the strong flares on these active stars threaten their habitability \citep[e.g.,][]{shields16,howard18,atri21}. The low stellar mass and the sharp habitable zone regime leave little dynamical room for scaled-down versions of the Solar System around these stars, which is suggested by the low occurrence rate of giant planets around M-dwarfs in comparison the high population in G-dwarfs (see \citealt{sabotta21}). In addition, the derivation of stellar parameters in this stellar regime is more difficult \citep[e.g., ][]{passegger22}, which directly affects the determination of the planet properties (mass and radius).  There is also some discrepancy in the determination of the metallicity of these stars through different methods, hindering the inference of reliable planet-metallicity correlations. By contrast, M-dwarfs host twice as many planets in the $m_p<10$~Mearth{} regime than solar analogs and they are specially located on orbits closer than 10 days \citep{sabotta21}.

On the other side, G-dwarfs host their habitable zones at periods around one year \citep{kopparapu13}. This creates critical difficulties in the detection of rocky planets through the radial velocity technique, requiring several-year-long campaigns with stable instruments at the centimetre-per-second level. So far, only ESPRESSO \citep{pepe20} is capable of doing this task and it will still be challenging for this ultra-stable instrument. Solar-like stars also have higher granulation levels and noise, thus increasing the stellar jitter \citep[see, e.g.,][]{dumusque11,cegla19,dravins21}. 

A new perspective is thus needed to face the detection of habitable worlds that can be confirmed and characterized by the radial velocity technique and provide a census of confirmed habitable rocky worlds with firm possibilities to host living organisms. In this paper, we present the KOBE (K-dwarfs Orbited By habitable Exoplanets) experiment\footnote{\url{http://kobe.caha.es}}, a legacy program of the Calar Alto Observatory with the CARMENES instrument to look for new habitable-zone planets around K-dwarfs. The paper is organized as follows. In Sect.~\ref{sec:opportunity}, we describe the opportunity that this stellar type regime offers in the instrumental plane and the astrobiological context. In Sect.~\ref{sec:KOBE}, we describe the goals and strategy of the programme based on the awarded observing time. In Sect.~\ref{sec:targets} we describe the target selection process followed to reach the final target list for the survey. Section~\ref{sec:observations} summarizes the supporting observations used during the selection process. In Sect.~\ref{sec:charact}, we present a preliminary characterization of the KOBE sample of targets based on the first year of spectroscopic observations. We present our conclusions in Sect.~\ref{sec:Conclusions}.

\section{The K-dwarf opportunity}
\label{sec:opportunity}

K-dwarfs, and more specifically, late K-dwarfs (K4-M0, with effective temperatures between 3800-4600 K) offer a compromise between technical and physical feasibilities to search for planets in the habitable zone. Several aspect are of key importance in this regime.

\subsection{Habitability and life detectability of K-dwarfs}
Although the conditions for habitability on the surface of a planet are still poorly understood, there are different properties that we know might hazard sustainable life as we know it. The first of these conditions is the ability of the planet to retain liquid water on its surface. The range of distances from the star where the incident flux on the planet allows this is called the habitable zone (HZ). 
In the case of mid-to-late M-dwarfs, given their low luminosity, this region is located at periods between 7-40 days. This, although favourable to the planet detection \citep{kaltenegger09}, has some key downsides on the real habitability of the planet. At first, stellar activity on M-dwarfs is a key actor \citep[e.g.,][]{jeffers18}, with energetic stellar flares increasing the luminosity of the star by a relevant percentage and increasing the coronal emission \citep[e.g.,][]{gunther20b}. Flares can also potentially reach the location of the habitable zone threatening any kind of life on its surface \citep[e.g.,][]{segura03,atri17}. Big stellar spots may also create relevant variations in the incident flux. Also, even the least active M-dwarfs have shown a significant X-ray emission, as well as UV (XUV) \citep{france13}. The great XUV emission is specially important regarding the true habitability of a planet \citep{shields16}, since it can cause atmosphere erosion \citep{lammer07}, as well as hydrodynamic scape \citep{luger15}. {These levels of XUV can also damage the cells structures of living organisms and strongly limit the proliferation of life on the surface of such planets}. Additionally, being closer to their parent stars, planets in the HZ might be tidally locked \citep{heller11}, always facing the same side to the parent star and thus decreasing the probability for life to be sustained in its surface \citep{barnes17}.

By contrast, K-dwarfs have their HZ located at longer periods (typically 50-200 days), where planets can have their rotation and orbital periods decoupled, thus allowing the planet to have day-night cycles. Stellar activity and magnetic flaring is dramatically diminished for stars earlier than M3 and specially in the late K-type domain \citep[see, e.g., ][]{hilton10,davenport16,gunther20b,astudillo-defru17}. Consequently, habitability is not threatened by these effects as much as it is in the HZ planets around M dwarfs. Besides, unlike in M-dwarfs, {we can derive, in a standard way, precise and reliable} stellar parameters, as well as chemical abundances \citep[e.g., ][]{sousa18} that are relevant to a proper characterization of the planets and the star-planet connection (although specific methodologies recently developed allow for a precise characterization of M-dwarfs, see e.g., \citealt{adibekyan18,Passegger2018,Marfil2021,abia20,shan21}). In this sense, the NIR spectra from CARMENES is an opportunity to get better abundances since at late K-dwarf temperatures, the optical spectra starts to contain significant molecules blending with atomic lines, which hampers the correct determination of atomic abundances. Ultraviolet radiation is also advantageous in K-dwarfs. Since they are less active and have shorter pre-main-sequence phases \citep{luger15}, the energetic radiation from near-UV to X-ray wavelengths is 5-50 times smaller at young ages than in the case of early M-dwarfs \citep{Richey-Yowell19}.

Overall, K-dwarfs are the Goldilocks stars that are prime targets in the search for life beyond Earth \citep{cuntz16}. If habitable, K-dwarfs also represent the best trade-off to detect biosignature molecules through direct imaging (e.g., through envisioned space-based interferometers, i.e., the LIFE mission\footnote{\url{https://life-space-mission.com}}, \citealt{quanz21}). {A planet within the HZ of a K7 dwarf at 30 pc has a maximum projected separation of 10 milli-arcsec, feasible with the next generation of space-based interferometers}. For example, \cite{arney19} demonstrated that  K-dwarfs offer a longer photochemical lifetime of methane in the presence of oxygen (this being considered an ideal biosignature) compared to G dwarfs. Looking for habitable planets around these stars (even if not transiting) is thus key for future searches of biosignatures through future proposed direct imaging space telescopes such as the LIFE mission \citep{quanz21}.

\subsection{Planet detections}

The K-dwarf habitable zone ranges from 0.4-1.2 AU for the earliest types to 0.1-0.3 AU for the latest  (corresponding to orbital periods between 17-200 days). This corresponds to radial velocity semi-amplitudes of 2.4 - 4.2 \ms{} in the case of M0 stars and 1-2 \ms{} for K4 stars with a 10 \Mearth{} planet in the habitable zone. 
Added to this, the imprint of stellar activity and magnetic cycles on the radial velocity of K-dwarfs is smaller than in the case of M-dwarfs \citep{santos10,lovis11}. For the late K-dwarf stars, the signals induced by magnetic cycles typically have amplitudes below 3 \ms{} and typical periods of 7 years. On the other hand, although rotation periods span between 15-45 days \citep{mcquillan14}, K-dwarfs in the colour range $1.0 < B - V < 1.3$ have the lowest level of activity jitter \citep{isaacson10},  significantly less than 1 \ms{}, thus becoming the perfect targets to search for habitable planets. 

With a demonstrated precision of around 1.3 \ms{} in the long-term (after corrections are applied, see \citealt{tal-or18}), CARMENES is one of the few instrument in the northern hemisphere that can reach such precision and stability over a long period of time. The 1.3 \ms{} precision allows the detection of planets in the HZ regime of K-dwarfs down to the rocky regime. Indeed, it allows for completeness with regard to planets with masses above 10 \Mearth{} around late K-dwarf stars and detection limits down to 3 \Mearth{} (see Sect.~\ref{sec:strategy}).

\subsection{The K-dwarf habitable-zone desert}
\begin{figure*}
\centering
\includegraphics[width=1\textwidth]{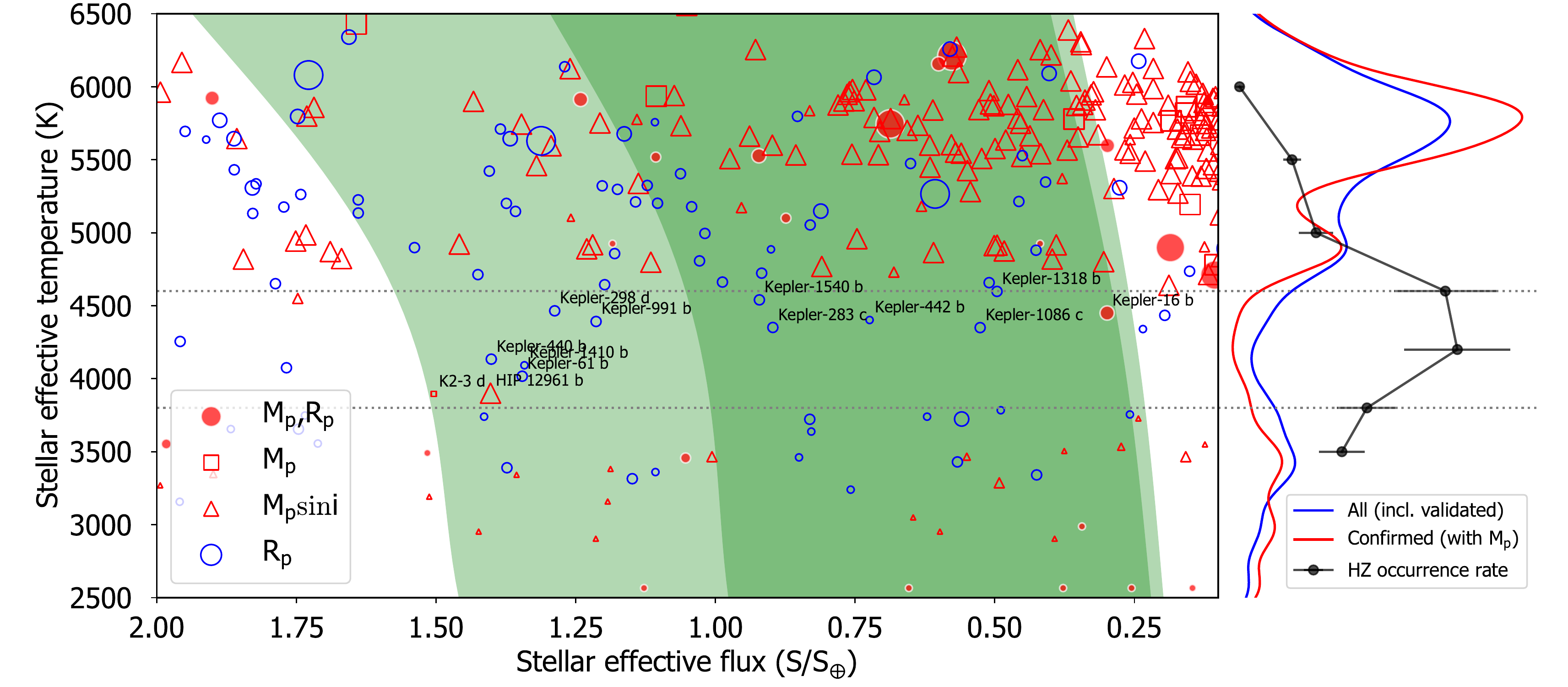}
\caption{{Detected extrasolar planets within the habitable zone for different stellar types}. The symbol code is shown in the legend and their size scales with planet radius for circle symbols and planet mass for squares and triangle symbols. Green regions show the optimistic and conservative habitable zones from the \cite{kopparapu13} climate models. The kernel density estimation (KDEs) on the right panel show the distribution of planets inside the HZ. {The black symbols and line correspond to the occurrence rates of HZ planets for the different stellar types obtained by convolving the \cite{kunimoto20} occurrence rates with the HZ parameter space. The two dotted horizontal lines encompass the late K-dwarf regime. The difference between the observational density profile and the expected occurrence clearly demonstrates the observational bias nature of this desert being due to a lack of focus on this regime.}}
\label{fig:hz}
\end{figure*}

The strategy followed by ground-based surveys and space-based missions has missed the HZ of K-dwarfs and, especially, late K-dwarfs. This is evident in Fig.~\ref{fig:hz}, where only four validated transiting planets (i.e., no mass measurement: Kepler-283 c and Kepler-298 d from \citealt{rowe14}; Kepler-440 b and Kepler-442 b from \citealt{torres15}) and one confirmed planet (HIP 12961 b, \citealt{forveille11}) populate the habitable zone in this stellar regime. The main properties of these few exoplanets are shown in Table~\ref{tab:hz}. The kernel density estimation (KDE) in Fig.~\ref{fig:hz} (right panel) illustrates this desert, with a large number of temperate worlds detected around G-type stars (mainly with RV surveys) and another large sample in the low stellar mass regime. This desert is even drier when we focus on planets with determined masses (red histogram). Stellar population studies, however, do not dispaly a paucity of this type of stars in the solar neighborhood, as compared to G- and M-types (e.g., Kroupa et al., 1993). Given that there is no theoretical evidence for a paucity of rocky planets in the HZ of K dwarfs, the origin of this desert is likely caused by the observational strategies applied in each approach.

\begin{table*}
\setlength{\extrarowheight}{4pt}
\centering
\caption{Main properties of known planets (either confirmed or validated) within the optimistic habitable zone around late K-dwarfs ($3800~K~<~T_{\rm eff}~<~4600$~K) as of May 2022. These planets are shown in Fig.~\ref{fig:hz}.}
\begin{tabular}{ccccccccc}
\hline\hline
Planet & Period & R$_{\rm p}$ & M$_{\rm p}$  & G$_{\rm mag}$ & T$_{\rm eff}$ & Discovery ref. \\
       & [days] & [\Rearth{}]  & [\Mearth{}] & [mag] & [K] &  \\
\hline

HIP 12961 b & 57.435 & -   & 114$\pm$22 & 9.6 & 3901 & \citet{forveille11} \\
K2-3 d & 44.55646 & - &  6.48$\pm$0.93 & 11.4785 & 3896 & \citet{crossfield15}\\
Kepler-1086 c & 161.5163345 & 2.94$\pm$0.11 & - & 15.56 & 4350 & \citet{morton16} \\
Kepler-1318 b & 213.257663 & 3.11$\pm$0.17 & - & 15.24 & 4598 & \citet{morton16}\\
Kepler-1410 b & 60.866168 & 1.78$\pm$0.12 & - & 15.95 & 4092 & \citet{morton16}\\
Kepler-1540 b & 125.4131177 & 2.49$\pm$0.19 & - & 14.08 & 4540 & \citet{morton16} \\
Kepler-283 c & 92.743711 & 1.82$\pm$0.12 & - & 15.88 & 4351 & \citet{rowe14} \\
Kepler-298 d & 77.473633 & 2.5$\pm$0.2 & - & 15.42 & 4465  & \citet{rowe14} \\
Kepler-440 b & 101.11141 & 1.86$\pm$0.24 & - & 15.11 & 4134 &  \citet{torres15}\\
Kepler-442 b & 112.3053 & 1.34$\pm$0.11 & - & 14.93 & 4402 &  \citet{torres15}\\
Kepler-61 b & 59.87756 & 2.15$\pm$0.13 & - & 15.04 & 4017 & \citet{ballard13} \\
Kepler-991 b & 82.5342519 & 2.54$\pm$0.13 & -  & 15.15 & 4392 & \citet{morton16}\\

\hline
\end{tabular}
\label{tab:hz}
\end{table*}

The reasons for this paucity of planets in the habitable zone are clear: the focus on solar-like stars for establishing a similarity with the Solar System and the hunt for planets around M-dwarfs due to reasons of detectability. A focused and systematic program exploring the habitable zone of K-dwarfs is thus missing. The M2K project \citep{apps10} in 2010 used a small fraction of the Keck/HIRES instrument to look for planets around MK stellar types, but only a handful of systems were announced. The HARPS GTO program (PI: X. Bonfils) has also followed-up several K-dwarfs \citep{bonfils05}. However, these studies were not focused on the HZ of K-dwarfs and the sample and cadence were thus insufficient to reach the rocky regime in the habitable zone on a significant number of targets. This was mainly due to the lack of telescope time. Hence, a dedicated service program with the flexibility of a moderate number of nights per semester and distributed along a sufficiently large time span would be required to reach this regime.

\section{The KOBE experiment: goals and observational strategy}
\label{sec:KOBE}

\subsection{Goals}
\label{sec:goals}
KOBE is the largest systematic and dedicated survey to search for habitable planets around late type K-dwarfs to date. The call for Legacy Programs from the Calar Alto Observatory\footnote{See the call in \href{https://www.caha.es/news/releases-mainmenu-163/13545-public-surveys-and-new-instrumentation-for-calar-alto-observatory}{this URL}.} was a unique opportunity to carry out a survey that would otherwise be practically impossible to be developed in any other facility (the chances of getting > 20 nights/semester in open time proposals on state-of-the-art instruments are extremely low, if not impossible). With the KOBE experiment we propose a guided search for habitable planets (from gaseous to rocky compositions) around a minimum of 50 late K-dwarfs by monitoring a carefully selected sample of K4-M0 stars. Our aim was to obtain an average of 90 radial velocity points per target spread over five semesters (see Sect.~\ref{sec:strategy}), using 35 nights per semester (including overheads) of the Calar Alto 3.5m telescope. {This time span and the proposed cadence described in subsequent sections would allow us to detect {and confirm} planetary signals with periods up to 300 days}. The experiment does not exclude gaseous giants in the habitable zone since this niche is also very relevant in different aspects (e.g., future search for habitable exomoons as in \citealt{teachey18,kipping22}-, co-orbital worlds as in \citealt{lillo-box18a,lillo-box18b}, or atmospheric characterization if they transit). 

Based on the planet occurrence rates from the Kepler mission (e.g., \citealt{kunimoto20}) and while taking into account the guided nature of this experiment, we expected to detect 15-40 new planets, with a relatively high percentage of them residing in the habitable zone and being in the super-Earth regime ($3<Mp<10$ \Mearth{}), detectable with CARMENES (see Sect.~\ref{sec:yield} on the expected yield of the program).

\subsection{Observing strategy and detectability rates}
\label{sec:strategy}
\begin{figure*}
\centering
\includegraphics[width=1\textwidth]{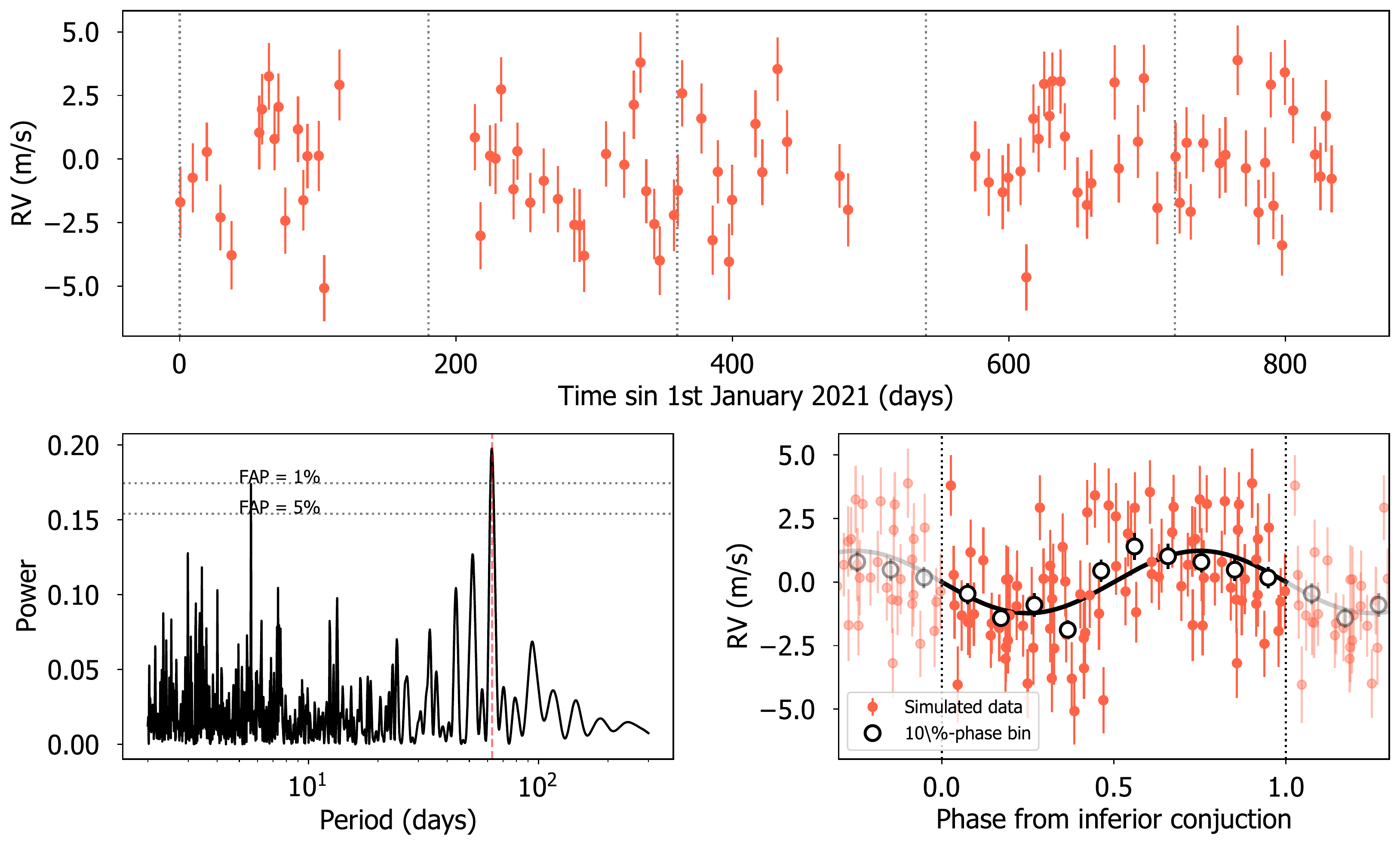}
\caption{Simulation of CARMENES data from a K7 star with a 5 \Mearth{} planet, with an 62.85-day orbital period using a strategy of 23 data points per semester over 4 semesters (middle panel in Fig.~\ref{fig:detectability}). Upper panel: Simulated radial velocity time series. Bottom left: Periodogram and the significance levels. The vertical dashed line shows the simulated period. Bottom right: Phase-folded radial velocity.}
\label{fig:simulation}
\end{figure*}

\begin{figure}
\centering
\includegraphics[width=0.5\textwidth]{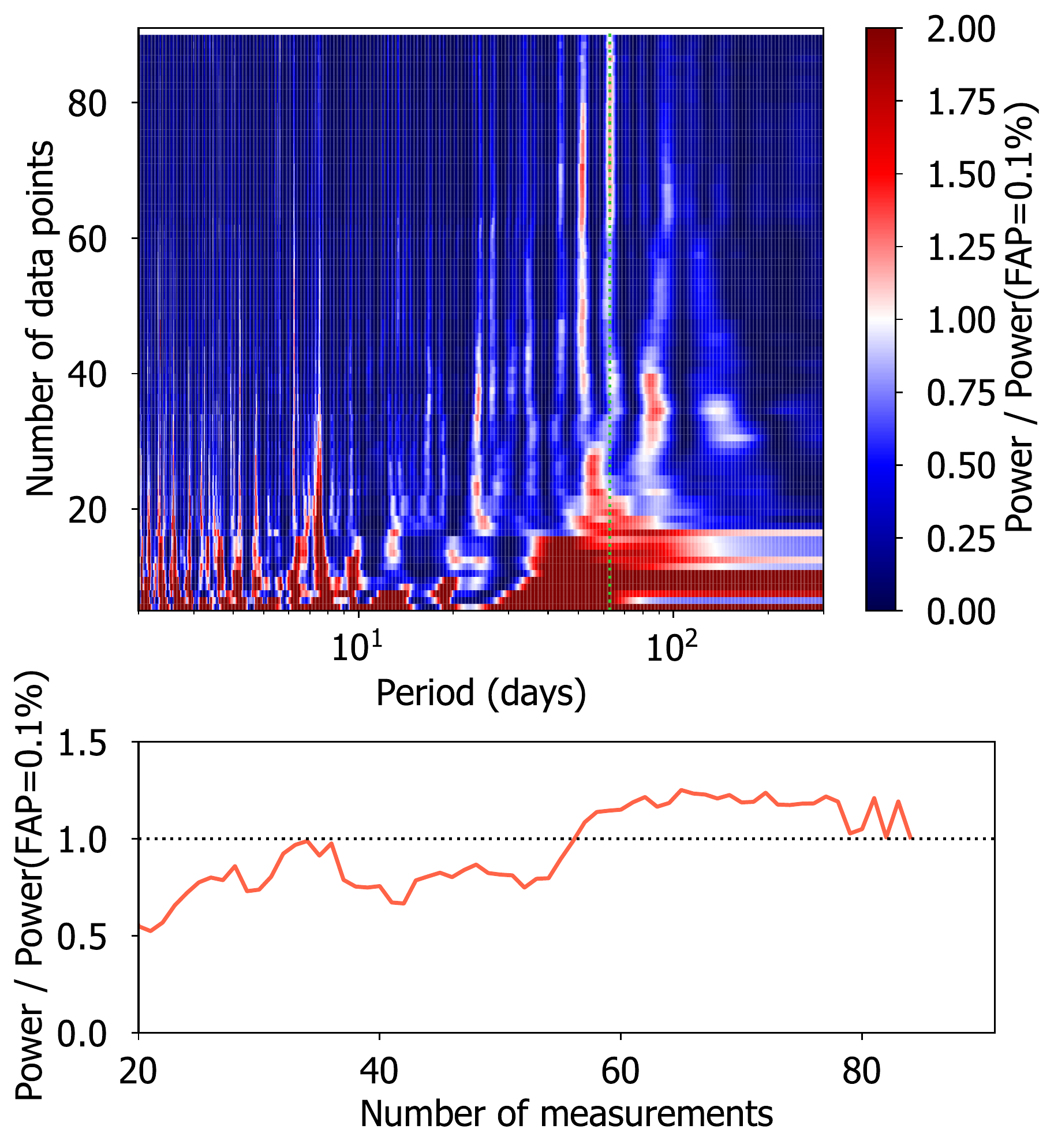}
\caption{Evolution of the GLS periodogram of the simulated RV dataset described in Sect.~\ref{sec:strategy} with the number of measurements and normalized to the FAP=0.1\% (upper panel), and specifically of the power at the planet period (lower panel).}
\label{fig:periodogram}
\end{figure}

As mentioned above, the ultimate goal of KOBE is to detect a significant sample of new habitable planets around late K-dwarfs. To that end, KOBE will sample a minimum of 50 targets. In order to estimate the necessary amount of measurements per target, we performed simulations focused on a K7 star ($M_{\star}=0.53$~\Msun{}), surrounded by a single 5 \Mearth{} planet located at an orbital period of $P_{\rm orb}=62.85$~days. We assumed a stellar jitter of 2 \ms{}, which includes the potential unaccounted effects of stellar activity. We assumed any activity above that level can be modeled by using Gaussian Process techniques commonly applied in planet detection (e.g., \citealt{suarez-mascareno20, faria22, lillo-box20,damasso20,demangeon21}). Based on current data, we assume a 1.3 \ms{} mean uncertainty on the individual CARMENES measurements and a time span of five semesters with an average cadence of one datapoint every four days\footnote{The actual distribution for the simulated dates of observation is a random number {drawn from a uniform distribution in the range from 2 to 6 days.}}. We simulate survey allocation and bad weather conditions by assuming that 70\% of the observing nights allow for proper observations and that 80\% of the nights in a given semester have a fraction of the night allocated to the program. We also assume that the target is only visible along 9 months per year, which is the average visibility window of the actual KOBE targets (see Sect.~\ref{sec:targets}). Under this setup, we are able to obtain a total of around 90 measurements along the five semesters. The RV simulation is shown in Fig.~\ref{fig:simulation}. In this particular case, the simulated signal is clearly recovered with a false alarm probability (FAP) below 0.1\% at the end of the survey\footnote{{The other significant peak in this periodogram corresponds to a period of 5.2~days, which matches the first harmonic of the average observing cadence in this particular simulation.}}. Indeed, this threshold is crossed in a steady situation after $N_{\rm obs}=56$ measurements. This is shown in Fig.~\ref{fig:periodogram}, where the upper panel shows the evolution of the Generalized Lomb-Scargle (GLS) periodogram for the RV data as new measurements are taken. The lower panel shows the power of the periodogram (normalized at FAP = 0.1\%) at the true period $P_{\rm orb}=62.85$~days as new measurements are taken. 

\begin{figure*}
\centering
\includegraphics[width=0.33\textwidth]{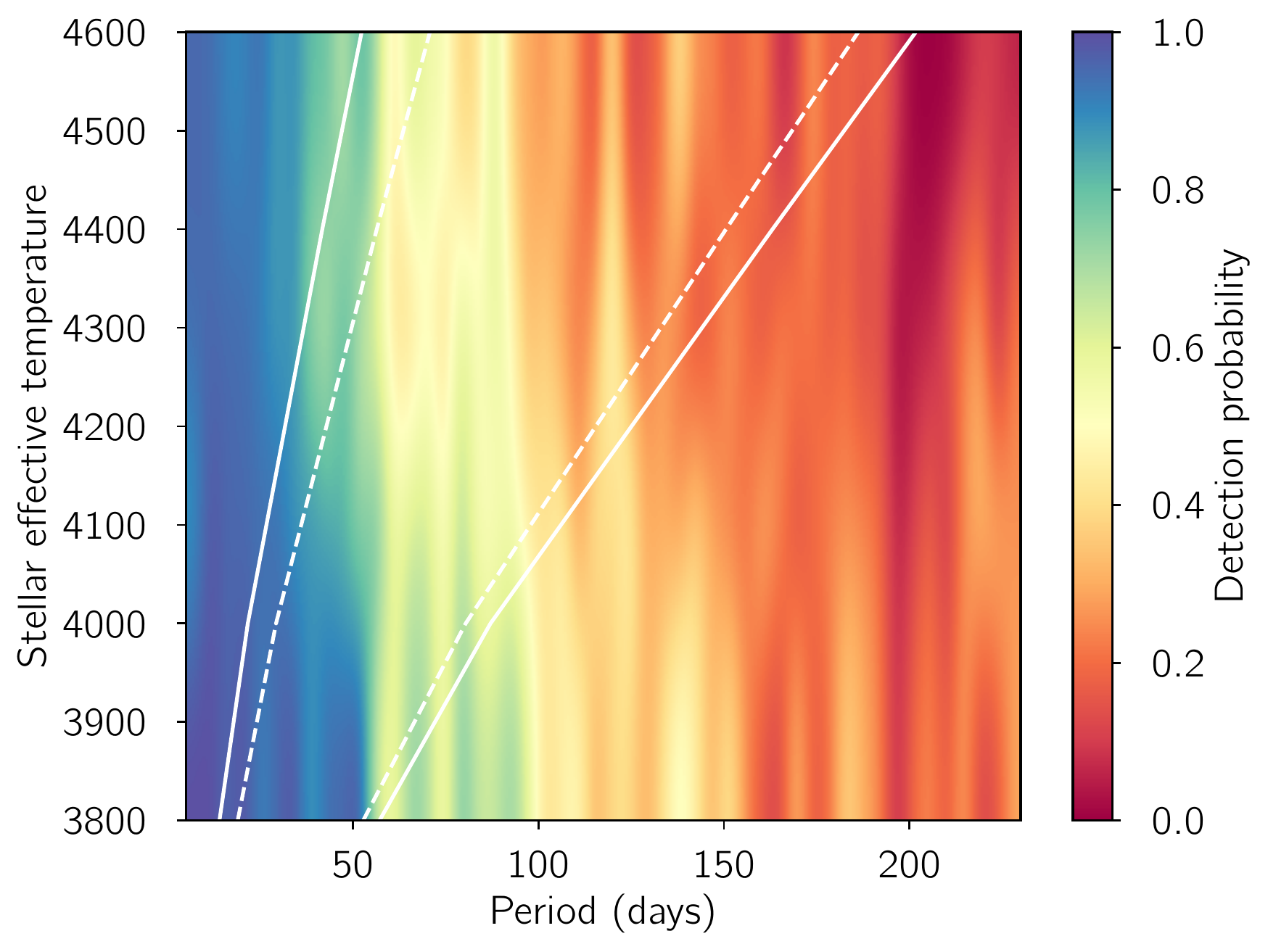}
\includegraphics[width=0.33\textwidth]{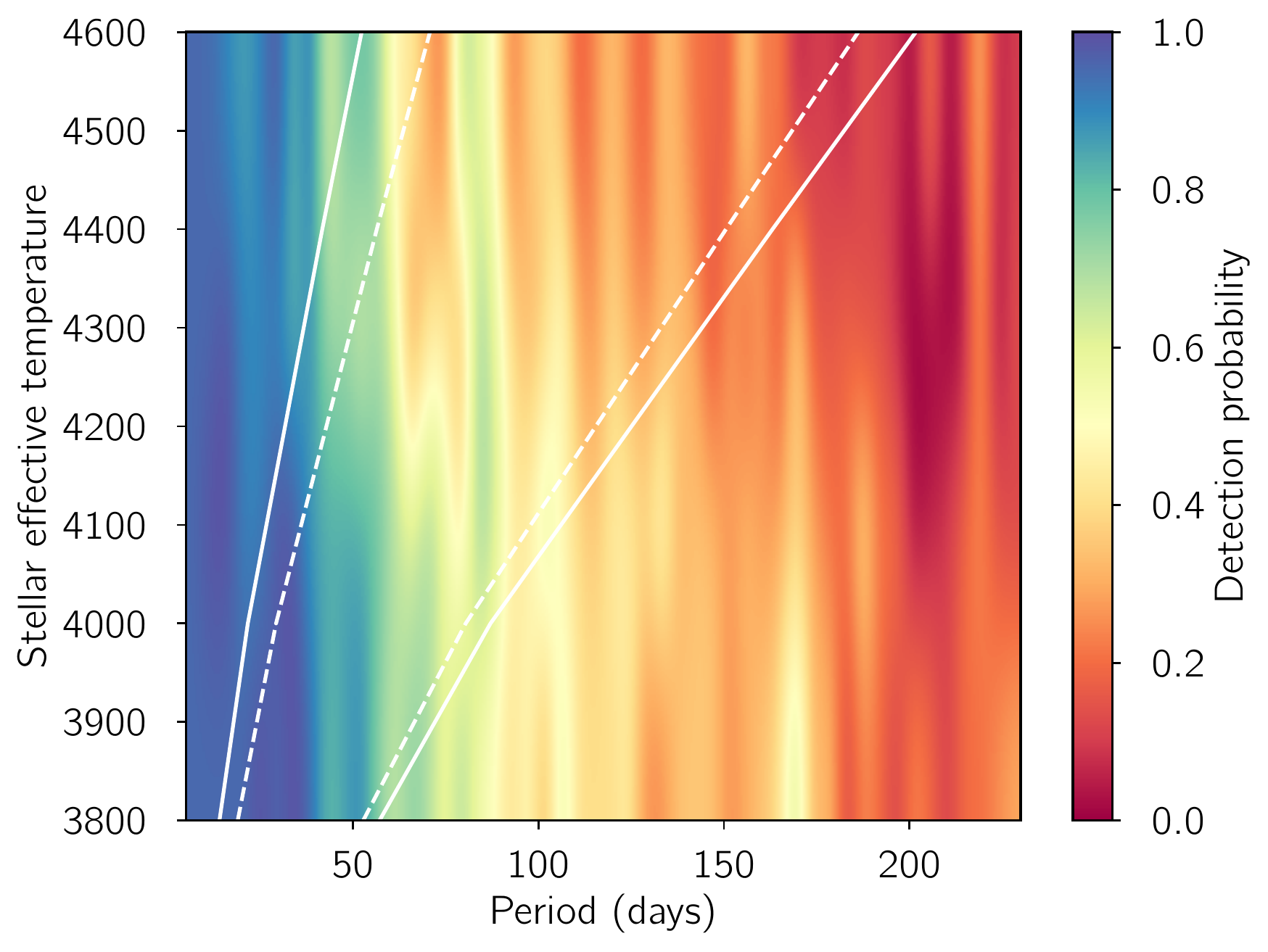}
\includegraphics[width=0.33\textwidth]{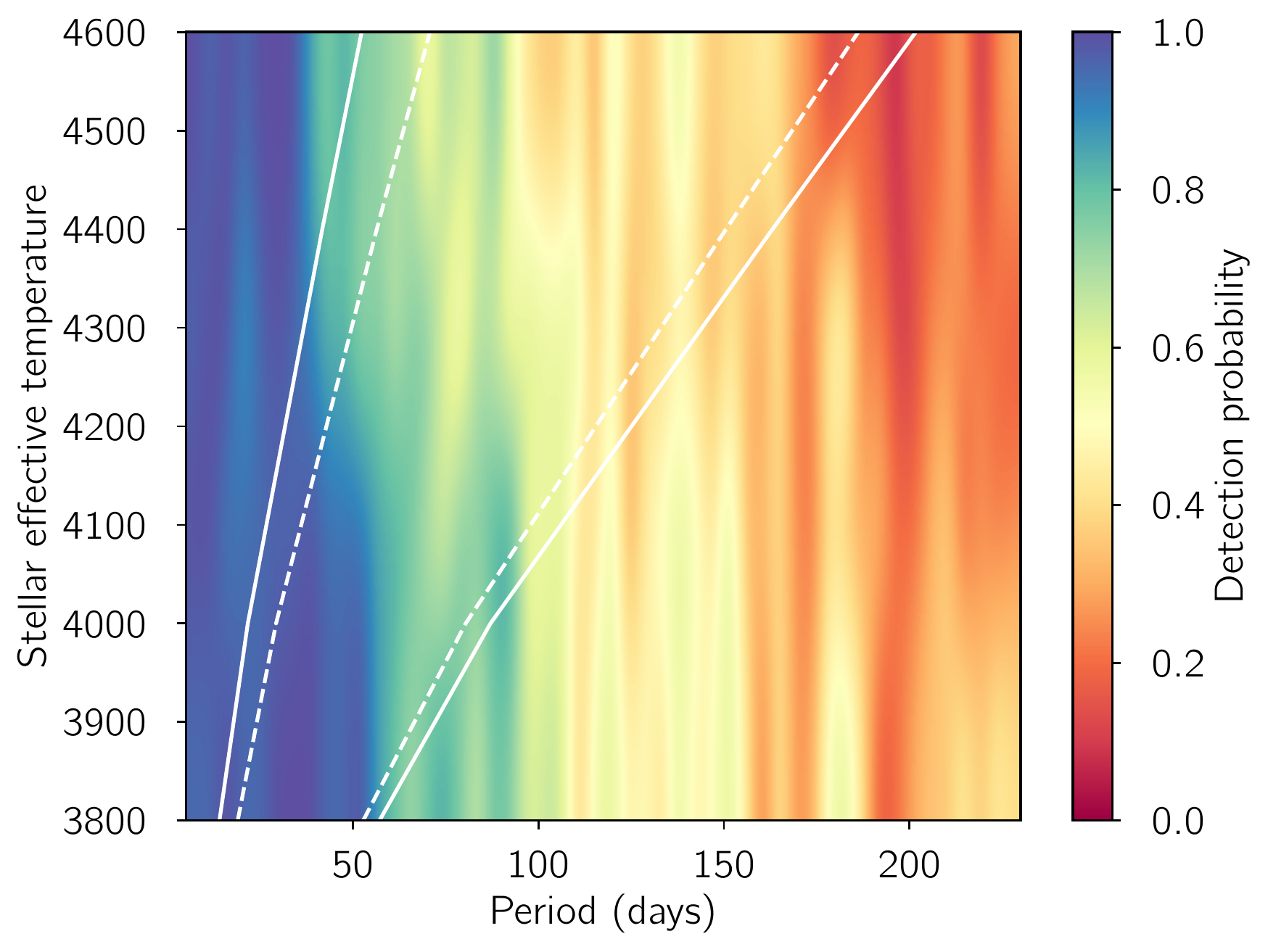}
\caption{Detectability maps of the KOBE experiment in the habitable zone ({white solid lines for the optimistic HZ and dashed for the conservative HZ}) of K4-M0 stars based on three observing strategies, from left to right: 4 semesters $\times$ 18 datapoints (left), $4\times 23$ (middle), and $5\times 23$ (right). Colour-code shows the detection probability for a 5 \Mearth{} planet at each location of the parameter space.}
\label{fig:detectability}
\end{figure*}

Although this is only one example, it illustrates the feasibility of the program to detect rocky worlds in the HZ of these late K-type stars in a moderate number of measurements. Based on this example, we now aim at estimating the detectability rates of the program based on different observational strategies. We designed three different strategies, combining different number of semesters (4 or 5) and data points per semester (18 - corresponding to a 6-day cadence-,  or 23 - corresponding to a 4-day cadence; both including the effect of bad weather and observability of the target), depending on the spectral type of the host star. We then assumed the following three strategies: 4 semesters with 18 datapoints per semester ($4\times 18$, 72 datapoints in total), $4\times 23$ (92 datapoints in total), or $5\times 23$ (115 datapoints in total). For each of the strategies, we created a grid of periods (from 5 to 230 days, {covering the HZ of all spectral types considered, i.e., K4-M0}) and stellar effective temperatures (corresponding to stellar masses from 0.49 to 0.61 \Msun{}). For each pair \{$P_{\rm orb}$, $M_{\star}$\} we simulated 20 RV time series including a Keplerian signal corresponding to a 5~\Mearth{} planet in a circular orbit and with the same observing setup and white noise and RV uncertainty as described in the example above. We then obtained the number of simulated time series where the injected periodicity ends up having a ${\rm FAP} < 5\%$. Figure~\ref{fig:detectability} (used as a detection limit corresponding to $2-\sigma$) shows the result of these detectability maps for each of the three observational strategies. 

 The results show that with the proposed strategy, we can have a detection probability for a 5 \Mearth{} planet in the middle of the HZ  >50\% for K4-M0 stars (i.e., >90\% for K7-M0, >70\% for K5-K7, and > 50\% for K4-K5). This exercise allows us to define the a priori observational strategy of the survey. We will use the $4\times 18$ strategy for K7-M0 targets, $4\times 23$ for K5-K7 targets, and the $5\times 23$ for K4-K5 targets. This strategy will ensure a minimum detection probability of 50\% of HZ planets down to 5\Mearth{} over the whole spectral type regime. 

Depending on the activity level of each star (usually below 1~\ms{}, which    we can estimate based on analysis of archival data), we adapt the observing strategy (cadence, total time span and number of measurements per night) to mitigate red noise and to properly sample the phases of known planets (see accompanying paper on the \texttt{KOBEsim} tool for planning observations by \citealt{balsalobre-ruza22}in).

\subsection{Expected yield of planets}
\label{sec:yield}

FGK stars are known to harbor at least one planet \citep[e.g.,][]{kunimoto20}. \cite{burke15} measured the planet occurrence rates from the results of the Kepler mission (using data from quarters Q1-Q16) and provided an average occurrence rate of 77\% for rocky planets ($0.75<R_p<2.5$~\Rearth{}) around GK dwarfs ($4200K<T_{\rm eff}<6100$~K), with an allowed range between 30\%-200\%. \cite{kunimoto20} performed a more detailed analysis splitting the occurrence rates into the three spectral types F, G, and K. Their results show that for K-dwarfs, the marginalized planet occurrence rate is $2.56^{+0.29}_{-0.26}$ planets per star with orbital periods below 400 days, while decreasing to $1.84^{+0.15}_{-0.14}$ for orbital periods below 200 days, it is still significantly greater than for G-type (1.17 planets per star) and F-type (0.53 planets per star) stars. In particular, these authors found that small planets ($1<R_p<2.8$~\Rearth{}) around K-type stars are about twice more abundant than around G-type stars and five times more abundant than around F-type stars. 
This, however, does not apply for larger planets ($R_p>2.8$~\Rearth{}) where the occurrence rates of the three stellar type regimes flattens. 

We go on to use Table B.1 from \cite{kunimoto20} to address the expected yield of planets based on the CARMENES capabilities and the observational strategy proposed for KOBE. {The study presented in \cite{kunimoto20} shows the result for the whole K-dwarf spectral type domain. Since there is no individual study (to our knowledge) that specifically focuses on the latest regime of this spectral class, here we assume their K-dwarf occurrence rate holds for K4-M0 stars.} We assume that we will be able to detect a planet when the induced minimum semi-amplitude corresponds to $2\times K_{\rm min} = 5\sigma_{\rm RV}/\sqrt(N_{\rm obs}/10)$, where $\sigma_{\rm RV}$ is the expected uncertainty in the RV measurements (assumed here to be 1.3\ms{}) and $N_{\rm obs}$ is the total number of RV datapoints assumed to be homogeneously distributed along the orbital phase. In practice, this means that the minimum detectable RV amplitude ($2\times K_{\rm min}$) must be greater than five times the uncertainty of the bin corresponding to 10\% of the orbital phase. With  $N_{\rm obs}=90$ measurements, we have $K_{\rm min}=1.08$~\ms{}. By using this semi-amplitude and assuming the mass-radius relations from \cite{chen17}, we find that the occurrence rate of planets around K-dwarfs conditioned by the detectability of the KOBE strategy and CARMENES precision is $1.68\pm0.25$ {planetary signals detected} per star in the range 10 to 640 days. This means that KOBE will be able to find {the signal of} more than one planet per star. {We note that given the time span of the observations of the survey, planetary signals with periodicities above 300~days will require additional observations beyond the survey to be confirmed.} If we restrict ourselves to the habitable zone, we find an expected yield per spectral subtype of $0.53\pm0.11$ for M0 stars, $0.56\pm0.13$ for K7 stars, $0.94\pm0.22$ for K4-K5 stars. Hence, we expect to detect a planet in the habitable zone with the KOBE survey strategy in at least half of the sample. If we further restrict ourselves to the rocky regime in terms of mass ($M_p< 10$~\Mearth{}) and within the HZ, the expected yield would be $0.09\pm0.02$ for M0 stars, $0.08\pm0.02$ for K7 stars, $0.00\pm0.02$ for K4-K5 stars. Consequently, for 50 targets, we expect to find between one and five rocky-mass planets within the HZ with this strategy.

\section{KOBE targets selection}
\label{sec:targets}
In a sample-limited survey, target selection is a major step to achieve the final goals of the project. Here, we describe the process followed to select the sample of targets for the KOBE experiment, the so-called  blind-search target list (BTL). 

\subsection{Preliminary blind-search target list (preBTL)}
\label{sec:preBTL}
We begin with a preselection of the targets (pre-BTL) from simple archival and observability criteria based on their coordinates and the Gaia DR2 catalog \citep{gaia18}: 1) stars with effective temperatures between 3800-4600 K (late K-type stars); 2) stars with stellar radii below 0.8~\Rsun{} (to avoid K-giants); and 3) stars observable from Calar Alto Observatory for at least 4 months per year. This criteria left us with 6458 stars composing our preBTL.

\subsection{The extended BTL (eBTL)}
\label{sec:eBTL}
In a second step, we built a merit function to rank the targets from the pre-BTL based on the following criteria:

\paragraph{Brightness.} We prioritized bright targets  to be able to obtain optimal results in the maximum exposure time executable with CARMENES (1800\,s). In practice, this means restricting the sample to stars with $G<12$ mag. The built merit function is flat below $G<6$~mag because we do not integrate for less than 10 minutes to avoid intrinsic noise from the star. From $G=6$ to $G=12,$ the function decreases with a second order polynomial from 1 to 0. This parameter has the greatest weight, namely, 48\% of the total merit function.

\paragraph{Stellar photometric noise.} We used the available TESS data to constrain the variability level of the star, using it as a proxy for its activity level that may induce RV variations. We computed the ratio between the TESS amplitude of photometric variations ($A_{\rm TESS}$) against the intrinsic photometric noise (${\rm rms}_{\rm TESS}$), so $A_{\rm TESS} / {\rm rms}_{\rm TESS}$. We want this ratio to be as small as possible and potentially close to 1 in order to select the most quiet targets. The details on the observations used to evaluate this parameter of the merit function are provided in Sect.~\ref{sec:TESS}. This parameter has 24\% weight of the total merit function.

\paragraph{Visibility window.} We prioritize targets observable from Calar Alto as many days per year as possible. This translates into a linear function with the number of months a target is visible at least 1.5h/night above 40$^{\circ}$ in the CAHA sky. This parameter has 10\% weight of the total merit function.

\paragraph{Effective temperature.} We slightly prioritize targets in the middle of the effective temperature range, that is, K7 stars, as they define the KOBE experiment. Hence, we apply an inverted parabola with a merit value of 1 at K7 and 0 at K3 and M1 types. This parameter has 10\% weight of the total merit function.

\paragraph{Simbad classification.} We used the "object type" entry of the Simbad CDS catalogue to discard possible targets known to be variable. The following values for this parameter are set to 0: "**", "EB", "CV", "XB", "LX", "HX", "Al", "Bl", "SB", and "Er*". This parameter has 5\% weight of the total merit function.

\paragraph{Systemic RV} To avoid potential problems with the contamination of the Moon, we prioritized stars with systemic RVs outside of the $\pm$37~k\ms{} regime. This parameter has 2\% weight of the total merit function.

\paragraph{Moving groups} Since the age of stars in moving groups is well characterized and this might be a relevant parameter to understand the habitability of the planets, we award stars in moving groups with a small extra weight. This parameter has 1\% weight of the total merit function.\\

By using this criteria and the described shapes and weights for the different parameters, we ranked the pre-BTL and chose the first 200 targets in the list. This is what we call the extended BTL (eBTL).

\subsection{The reduced extended BTL (reBTL)}
\label{sec:reBTL}

For these 200 targets, we performed a more detailed inspection of the available data. We carried out the procedure using Spanish Virtual Observatory tools (SVO\footnote{\url{http://svo.cab.inta-csic.es}}) and by searching in archives of the main high-resolution spectrographs, such as SOPHIE, HARPS, or HARPS-N.

First of all, we ensured that the KOBE experiment is focused on the detection of new habitable worlds. Thus, we ruled out stars in which exoplanets have already been discovered or those highly monitored by other experiments.

Since planets are more commonly aligned with the stellar spin axis \citep{Albrecht21} and the RV technique is not sensitive to face-on systems, we also rejected stars with a null projection of their spin (i.e., with $\upsilon\sin{i_{\star}}<0.1$~k\ms{}, where $i_{\star}$ is the angle between the spin axis of the star and the plane containing the direction of sight.). Additionally, in order to obtain the best RV precision possible with the instrument, the absorption lines must be narrow. Stars of high rotation speed have very broad spectral lines making them less contrasted and overlapping, making it challenging to measure the Doppler shift produced by orbiting planets (e.g. \citealt{lovis11}). For this reason, we also avoided stars with $\upsilon\sin{i_{\star}}>8$~k\ms{}.
    
\begin{figure*}
\includegraphics[width=1\textwidth]{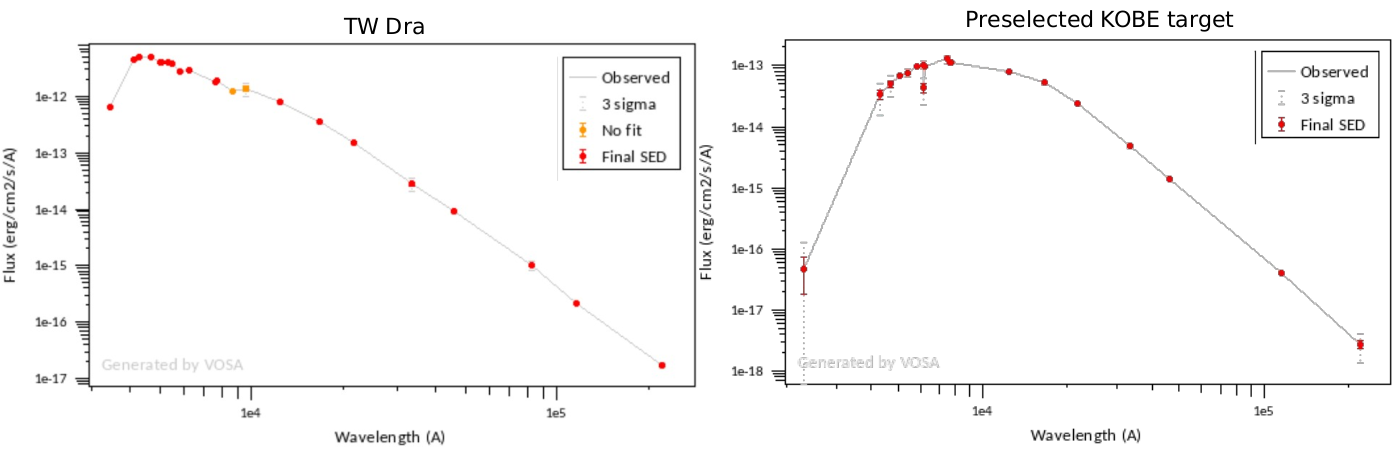}
  \caption{SED comparison of a binary system and an isolated star. \textbf{Left:} Binary system TW Dra of A-K type stars. \textbf{Right:} Pre-selected  K-dwarf for the KOBE sample. The SVO tool VOSA is used to obtain these graphs.}
  \label{fig:vosa}
\end{figure*}
    
Another constraint that we impose is the selection of single stars to avoid contamination from spectroscopic binaries. To verify the absence of companion stars to some extent, we studied the spectral energy distributions (SEDs) of the eBTL sample. By using the Virtual Observatory SED Analyzer (VOSA, \citealt{bayo08,rodrigo17}), we checked for SEDs showing different components, although with this we are only sensitive to companions with significantly different spectral types (details are given in Sect.~\ref{sec:SEDs}). In Fig.~\ref{fig:vosa}, we show the comparison of two SEDs from the eBTL sample. One of them corresponds to the binary system TW Dra (\citealt{Tkachenko10}), composed of two stars of types, A and K, where the two hot and cold components are evident. The other shows one of the KOBE candidate stars. In the former, an excess of flux at high frequencies is evident explained by the energy emitted by the A-type companion, while in the latter, there is apparently a black-body spectrum with any contamination. Targets showing the former behavior are discarded. 

In the search for possible binaries, we also check for close companions identified in the Gaia EDR3 catalog (Gaia Early Data Release 3, \citealt{gaia}) by using the \texttt{tpfplotter}\footnote{This code is publicly available in Github through the following URL:  \url{https://github.com/jlillo/tpfplotter}.} tool \citep{aller20}. For cases with identified companions with angular separations less than the size of the CARMENES fiber (1.5 arcsec), this is a compelling reason to discard the stars since the spectra would be contaminated by the objects in the field of view. We also checked the renormalized unit weight error (RUWE) value provided by the EDR3 catalog. According to the Gaia documentation, this parameter provides an estimate of the goodness of the single star fit to astrometric observations. For values larger than 1.4, this is an indication that either the star is not single or there is an issue with the astrometry. To be flexible, we discard sources with RUWE$>2$.
Based on this manual inspection, a total of 30 targets were discarded, leaving us with a reBTL composed of 170 stars. 

\subsection{Final selection: BTL and backup BTL (bBTL)}
\label{sec:BTL}

The final step in our selection process consists on a new merit function to rank the remaining stars in the reBTL. The new merit function consists on the following parameters:

\paragraph{Brightness.} We used the same function for this parameter as explained in the previous section. This parameter has the largest weight, namely, 25\%.

\paragraph{Contamination by close sources.} We used the Gaia EDR3 catalog to estimate the level of contamination inside the CARMENES fiber from close-by sources. We use a linear decay from 0\% to 0.1\% of contamination, providing null points to stars with contaminations larger than 0.1\%. This parameter has a weight of 25\%.

\paragraph{HZ coverage.} We prioritize targets where the observability time span over the course of the year is high compared to the HZ periods. This is to ensure that in a given observing season for that particular target, we can cover more than one orbit of planets within the HZ so that we are not strongly affected by the observing gap due to the target setting for some months. We take this into account in the merit function as second order polynomial so that stars where only half of the HZ period can be covered in an observing season is 0.1, while being 1 if three or more HZ periods fit in a single observing season. This parameter has a weight of 16.7\% in the merit function.

\paragraph{Stellar $\upsilon\sin{i}$.} We prioritized stars with low (although not null) values of the $\upsilon\sin{i}$ through a linear decay from 1 to 12~k\ms{}. The values from the $\upsilon\sin{i}$ are mostly obtained from a vetting campaign with CARMENES described in Sect.~\ref{sec:vetting} and "FWHM_vel_mean" parameter of the Gaia DR2 catalog when no vetting data was available. This parameter has a weight of 8.3\% in the merit function.

\paragraph{Homogeneous sky location.} The reBTL list contains an overabundance of targets at right ascensions (RA) from 15h-21h. In order to have enough targets over the course of the year, we slightly prioritize stars with RA coordinates outside of this range by giving them an extra weight of 8.3\%. \\

\paragraph{Effective temperature.} {We use the same weighting shape for the effective temperature as in previous steps (see Sect.~\ref{sec:eBTL}) to keep the preference for K7-type stars. The maximum weight is 8.3\% for this parameter.}  \\

\paragraph{X-ray emission.} {For this sample, we checked the X-ray emission from the ROSAT satellite with the aim of discarding potentially active sources in this regime. None of the targets was found to have a relevant emission. A weight of 8.3\% was assigned to this parameter.}  \\

\begin{figure*}
\includegraphics[width=1\textwidth]{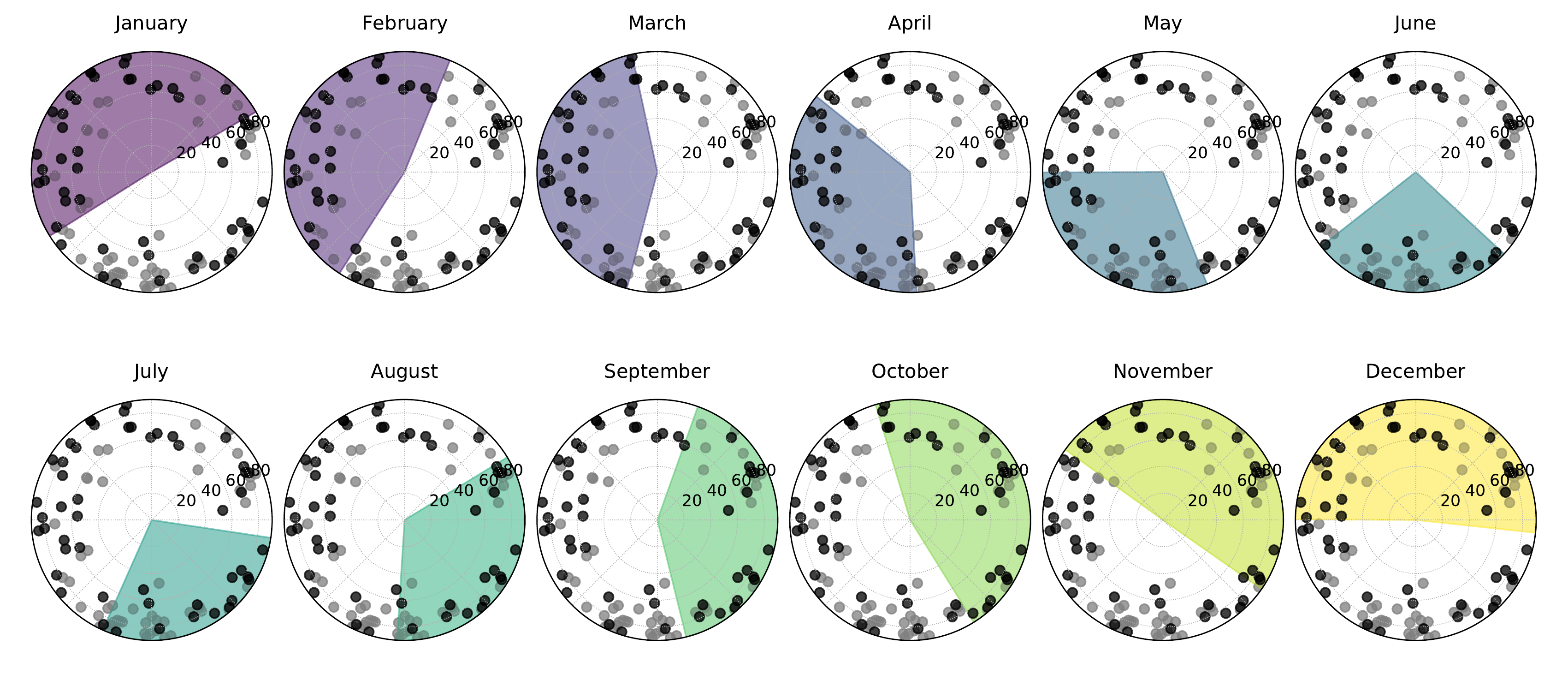}
  \caption{Location and visibility of the KOBE sample including the BLT (black symbols) and bBTL (gray symbols). For each month, the shaded colored region represents the region of the sky observable from CAHA.}
  \label{fig:BTLvisibility}
\end{figure*}

\begin{figure*}
\includegraphics[width=1\textwidth]{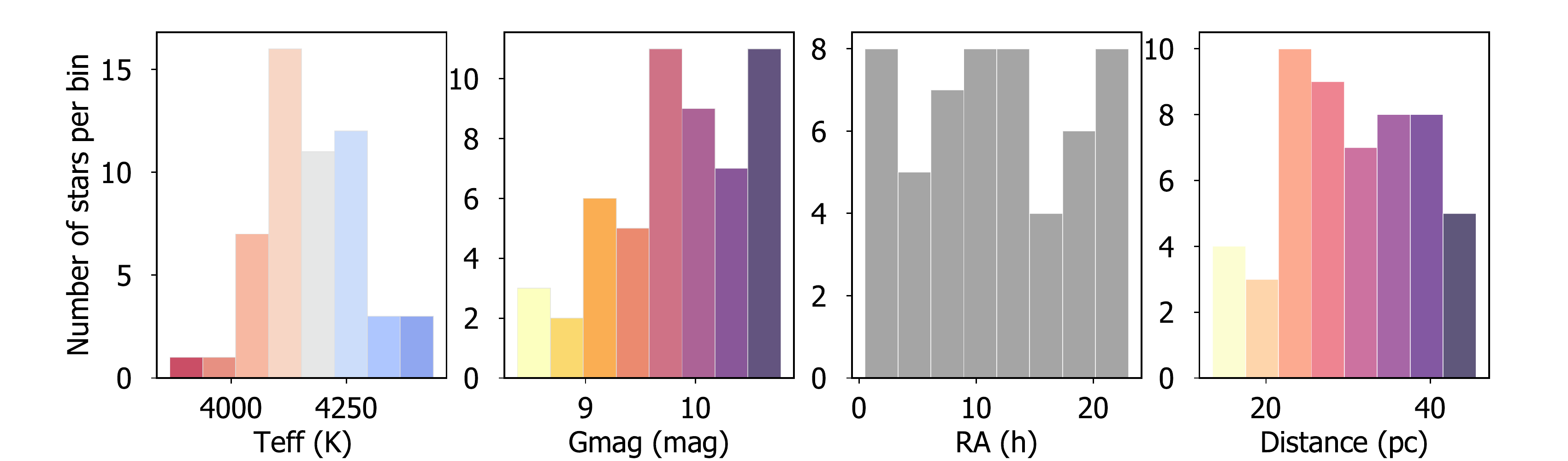}
  \caption{Properties of the final KOBE target list (BTL).}
  \label{fig:BTLprop}
\end{figure*}

The evaluation of the merit function described above into the reBTL targets produces a sorted list. From this list, we take the top 50 targets as the final BTL sample and the second 50 targets as the backup BTL (bBTL) sample. The visibility of the BTL and bBTL samples for each month from the CAHA observatory is shown in Fig.~\ref{fig:BTLvisibility}. And the general properties of this final sample (including effective temperature, Gaia magnitude, RA and distance) are summarized in Fig.~\ref{fig:BTLprop}.

\section{Supporting observations for target selection}
\label{sec:observations}

\subsection{TESS photometry}
\label{sec:TESS}

We used observations collected by the transiting exoplanet survey satellite (TESS, \citealt{ricker14}) during the first two years of its operation (sectors 1 to 26), to make an assessment of the photometric variability of the initial list of 6458 potential KOBE targets (the pre-BTL, see Sect.~\ref{sec:preBTL}).

We used the 2-min cadence lightcurves and the TESS-SPOC light curves from the full frame images provided by the TESS science team and available for download from the Mikulski Archive for Space Telescopes (MAST\footnote{\url{https://mast.stsci.edu/}}). In both cases, we used the PDCSAP (pre-search data conditioning simple aperture photometry) fluxes. Details of data processing steps included in the pipelines that produced these light curves can be found in \cite{jenkins16} and \cite{caldwell20}.
We searched the MAST archive for the light curves using the Gaia ID in order to avoid false identifications and analyzed 6404 2-min light curves for 2990 objects (many objects were observed in two or more sectors) and 3351 full frame images (FFIs) light curves for 1936 objects, of which only 230 did not have 2-min light curves. We analyzed each sector separately.

Prior to the variability analysis, data points marked as low quality in the TESS data files were excluded. Since the light curves also exhibit several types of instrumental artifacts (including slow drifts and significant deviations at the beginning or end of the continuous data segments), some light curve sections were manually removed. Once we had clean light curves, we automatically searched for variability signals by measuring the excess of noise in terms of the amplitude of the light curves over the root mean square (RMS) and the RMS using different bins (e.g., every 2 hours and every 10 hours). A Lomb-Scargle periodogram calculation was also performed in order to search for periodic signals.

We inspected a set of light curves by eye in order to set a threshold for those values (amplitude, rms, and amplitude/rms) to automatically discard variable sources. We also discarded eclipsing binaries and very low amplitude variables for which a periodicity could be found. In this way, we selected  1626 objects from the 2-min (PDCSAP) light curves and 940 objects from the FFI light curves.

\subsection{Spectral energy distributions}
\label{sec:SEDs}

\noindent To verify the absence of companion stars, we studied the spectral energy distributions (SEDs) to evaluate possible contamination by additional sources with significantly different spectral types. We obtained the distributions by means of \textit{VOSA}\footnote{\url{http://svo2.cab.inta-csic.es/theory/vosa/}} (\citealt{2008A&A...492..277B}), a spectral analyzer from the \textit{Spanish Virtual Observatory} (SVO\footnote{\url{{http://svo.cab.inta-csic.es}}}) tool. This facility provides a complete SED gathering archival photometric data in the whole electromagnetic spectrum range (e.g., from \textit{2MASS All-Sky Point Source Catalog} \citealt{2006AJ....131.1163S}, \textit{Tycho-2 Catalog} \citealt{2000A&A...355L..27H}, and \textit{Pan-Starrs PS1 DR2} \citealt{2002SPIE.4836..154K} among others) and fits it minimizing the reduced $\chi^2$. As our targets are K-late dwarfs we expect to have a black-body spectrum peaking around the visible and near-infrared range. Thus, we ruled out every star with any excess of flux at shorter (UV) or longer wavelengths (IR), as these would arguably be caused by the presence of a hotter or a colder companion, respectively (e.g., \citealt{2015BaltA..24..137C}).

\subsection{High-spatial resolution imaging}
\label{sec:astralux}

\begin{figure}
\includegraphics[width=0.48\textwidth]{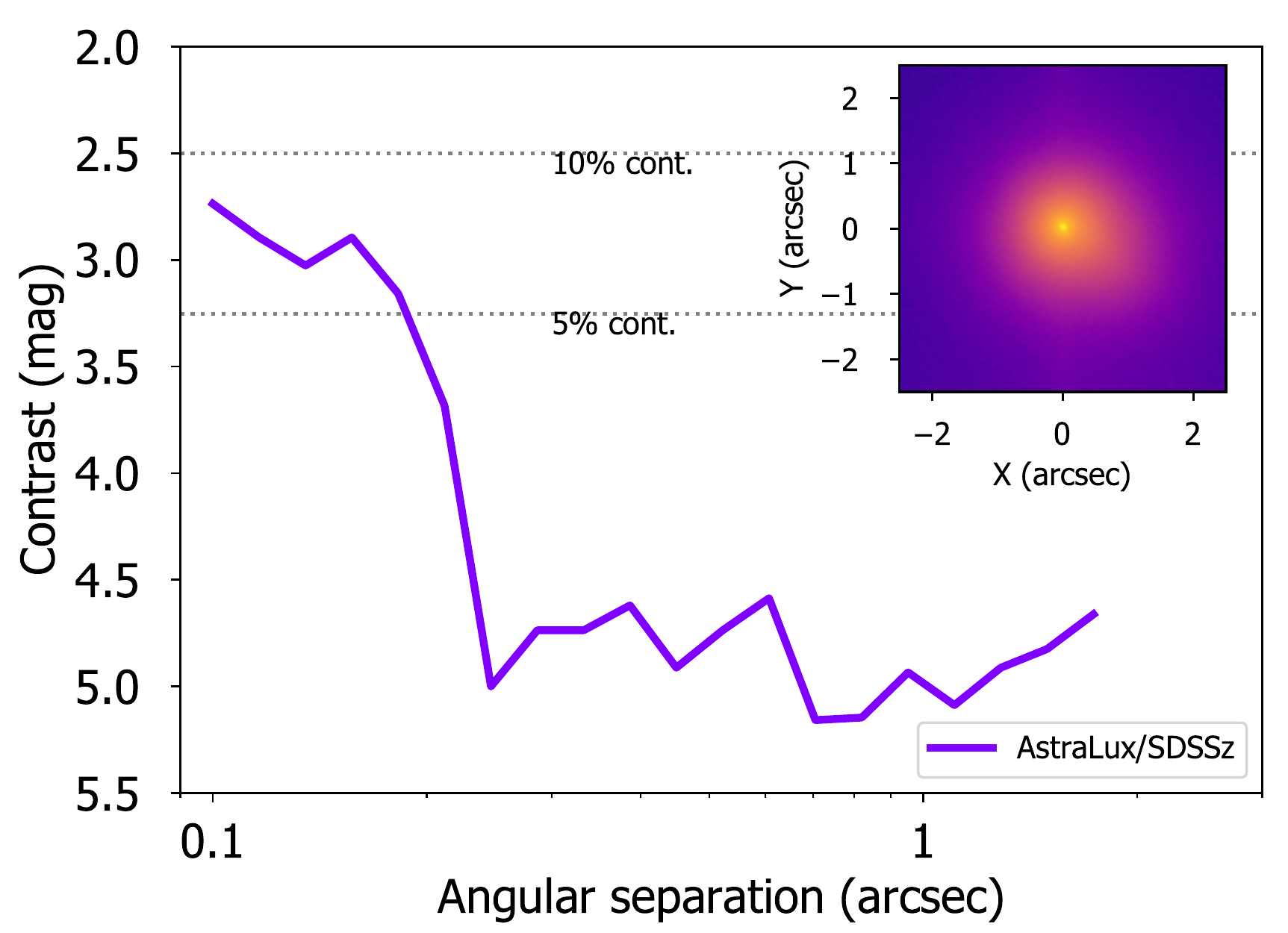}
  \caption{Example of the high-spatial resolution image (inset plot) and contrast curve (main plot) obtained with the AstraLux camera at the 2.2-meter telescope in Calar Alto Observatory for one of the BTL targets.}
  \label{fig:astralux}
\end{figure}

We obtained high-spatial resolution images for the BTL sample using the AstraLux instrument \citep{hormuth08} at the 2.2m telescope in Calar Alto Observatory. This instrument uses the lucky-imaging technique by performing fast readout imaging creating datacubes of thousands of short-exposure frames (below the coherence time) to subsequently select the ones with the highest Strehl ratio \citep{strehl1902} and combine them into a  final high-spatial resolution image. This process is done by the observatory pipeline \citep{hormuth08}. Using the Sloan Digital Sky Survey z filter (SDSSz), we observed the 50 targets of the BTL sample on 30 December, 2020 to further explore the possible contamination from sources closer than 1.5 arcsec. We used a $6\times6$ arcsec field of view allowing the individual exposure times to go down to 10 ms. We obtained between 2000 and 50\,000 frames depending on the target magnitudes and selected the 10\% best frames by using the Strehl ratio \citep{strehl1902}. We then proceeded to the alignment and stacking of the selected frames. Based on this final image,  we computed the sensitivity curve by using our own developed \texttt{astrasens} package\footnote{\url{https://github.com/jlillo/astrasens}} with the procedure described in \cite{lillo-box12,lillo-box14b}. We find no evidence of additional sources within this field for any of the observed targets and within the computed sensitivity limits. Overall, we can discard contaminant sources down to 0.2 arcsec with contrast magnitudes in the SDSSz band of brighter than $\Delta m_{\rm SDSSz} < 5$~mag. An example of this contrast curve and high-spatial resolution image is shown in Fig.~\ref{fig:astralux}.

\subsection{Vetting spectroscopy with CARMENES}
\label{sec:vetting}

We obtained one spectrum per target for all the BTL and bBTL samples to check that the rotational velocity of the stars in the sample are small enough as to obtain precise RV measurements. In total, all 100 stars were observed and all of them presented FWHM$<9$~k\ms{}. The observations were carried out in Decemeber 2020, prior to the start of the observations of the KOBE experiment. The data will be released together with the whole KOBE dataset at the end of the survey in 2024-2025.

\section{Spectroscopic characterization of the KOBE sample}
\label{sec:charact}

\begin{figure*}
\includegraphics[width=1\textwidth]{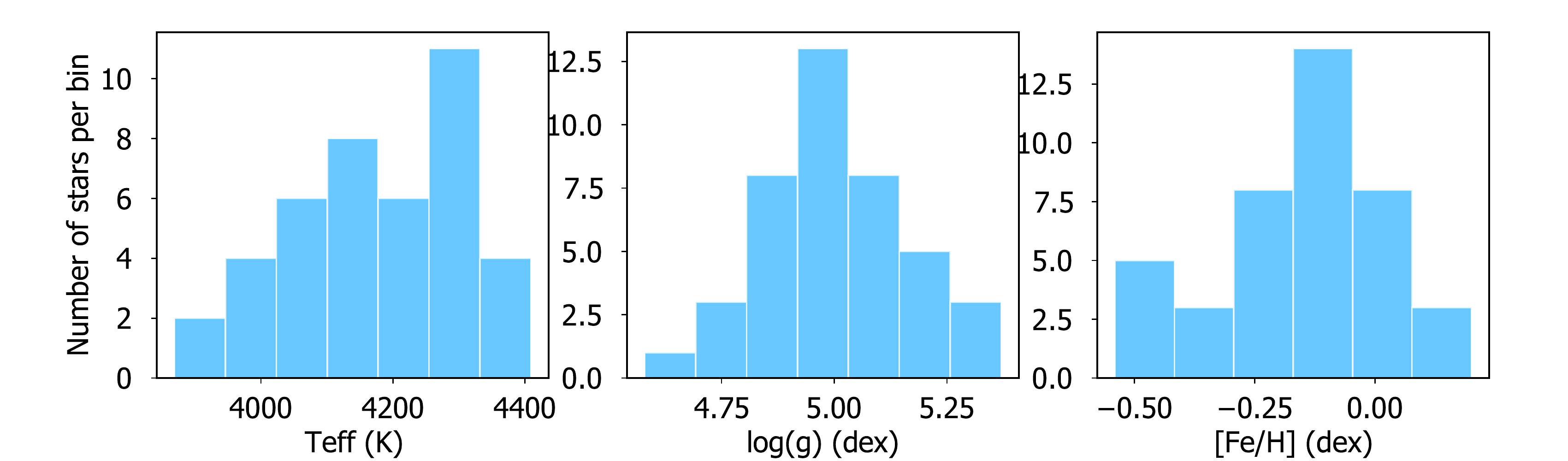}
  \caption{Spectroscopic properties of the final KOBE target list derived by using the averaged CARMENES spectra obtained along the first year of observations, as described in Sect.~\ref{sec:charact}.}
  \label{fig:stepar}
\end{figure*}

We used the first 10 months of KOBE data to create high signal-to-noise spectra of the KOBE targets in order to perform a detailed spectroscopic characterization of their properties. We adapted the method described in \citet{Marfil2021} for the analysis of the CARMENES M-dwarf sample to derive the stellar atmospheric parameters of the KOBE sample (i.e., effective temperature, $T_{\rm eff}$, surface gravity, $\log{g}$, and metallicity, [Fe/H]). We used the {\sc SteParSyn} code \citep{Tabernero2022} as a Bayesian implementation of spectral synthesis along with a synthetic grid based on BT-Settl models \citep{Allard2012}. The original grid in \citet{Marfil2021} was expanded towards the hot regime (up to 5000\,K) to effectively cover the parameter space encompassed by K-type dwarfs. However, we restricted the analysis to the CARMENES optical (VIS) channel, leaving 55 lines of \ion{Fe}{i} and \ion{Ti}{i} out of the 75 lines reported in \citet{Marfil2021}. We also excluded the TiO $\epsilon$ bands since their sensitivity to $T_{\rm eff}$ is considerably lower than the TiO $\gamma$ bands for hotter stars \citep{Marfil2021, Passegger2018}. We safely assumed a fixed $\upsilon\sin{i}=2$~km\,s$^{-1}$ since most targets exhibit no significant rotation, except for the star BD+11 514, for which we assumed $\upsilon\sin{i}=7.3$~km\,s$^{-1}$ \citep{MartinezArnaiz2010}. We compared our results with some other studies, including \citet{Lepine2013}, \citet{Gaidos2014}, \citet{Mann2015}, \citet{Boeche2016}, \citet{Deka2018}, and \citet{Morris2019}. The result on coincident targets show a good agreement on the properties derived through our methodology. Figure~\ref{fig:stepar} shows the histograms of the derived properties by using this methodology. The typical uncertainties for these properties is 50~K for the effective temperature, 0.25~dex for the surface gravity, and 0.14~dex for the metallicity. 

\section{Summary and conclusions}
\label{sec:Conclusions}

We have elaborated on the gap among habitable-zone planets around late-type K-dwarfs among the current population of extrasolar planets. Given the current estimates of planet occurrence, this desert is only due to observational biases, with most surveys focusing on G- and M-dwarfs for similarity to our own solar system and for detectability reasons, respectively. 

To fill this relevant parameter space in terms of astrobiological exploration (given the optimal properties of these stars to host habitable planets and the advantages of detecting planets in these regions around late K-dwarfs), we initiated the KOBE experiment, a radial velocity monitoring survey with CARMENES of 50 late-type K-dwarfs (K4-M0) in search for planets in their habitable zone. The survey is scheduled to run from 2021 to 2024. We carried out a detailed four-step target selection process (see Sect.~\ref{sec:targets}) to ensure the sample selection is optimal for planetary searches. This selection included data from different techniques, including TESS photometry, high-spatial resolution imaging, inspection of the \textit{Gaia} catalog, analysis of the spectral energy distributions, vetting spectroscopy, and inspection of the target properties including magnitude, sky coordinates and visibility. Our final sample has a roughly Gaussian distribution in effective temperature around 4100~K (with a slight overdensity around 4300K) and {a distribution of metallicities similar to that of the solar neighborhood (see \citealt{udry00,santos04}) ranging from -0.55 to +0.25 dex, with a median around [Fe/H]~$ = -0.07$~dex}. Based on the convolution of planet occurrence rates and detectability with our observing strategy, we estimate that KOBE will be able to find at least one planet within the HZ every two stars in the sample, providing an average number of around 25 new planets in the K-dwarf HZ desert. This will represent a major increase in the population of this very relevant regime in the search for life beyond Earth.

\begin{acknowledgements}
We thank J. Iglesias-P\'aramo for coordinating the observations of the KOBe Legacy from the Observatory side. J.L-B., O.B.-R. and A.C.-G. acknowledge financial support received from "la Caixa" Foundation (ID 100010434) and from the European Unions Horizon 2020 research and innovation programme under the Marie Slodowska-Curie grant agreement No 847648, with fellowship code LCF/BQ/PI20/11760023. This research has also been partly funded by the Spanish State Research Agency (AEI) Projects No.PID2019-107061GB-C61 and No. MDM-2017-0737 Unidad de Excelencia "Mar\'ia de Maeztu"- Centro de Astrobiolog\'ia (INTA-CSIC).
A.M.S acknowledges support from the Funda\c{c}\~ao para a Ci\^encia e a Tecnologia (FCT) through the Fellowship 2020.05387.BD. and POCH/FSE (EC). This work was supported by FCT - Funda\c{c}\~ao para a Ci\^encia e a Tecnologia through national funds and by FEDER through COMPETE2020 -  Programa Operacional Competitividade e Internacionaliza\c{c}\~ao by these grants: UID/FIS/04434/2019; UIDB/04434/2020; UIDP/04434/2020; PTDC/FIS-AST/32113/2017 \& POCI-01-0145-FEDER-032113; PTDC/FIS-AST/28953/2017 \& POCI-01-0145-FEDER-028953; PTDC/FIS-AST/28987/2017 \& POCI-01-0145-FEDER-028987.
We acknowledge the support by FCT - Funda\c{c}\~ao para a Ci\^encia e a Tecnologia through national funds and by FEDER through COMPETE2020 - Programa Operacional Competitividade e Internacionaliza\c{c}\~ao by these grants: UID/FIS/04434/2019; UIDB/04434/2020; UIDP/04434/2020; PTDC/FIS-AST/32113/2017 \& POCI-01-0145-FEDER-032113; PTDC/FISAST/28953/2017 \& POCI-01-0145-FEDER-028953.
The French group acknowledges financial support from the French Programme National de Plan\'etologie (PNP, INSU).
\end{acknowledgements}

%
%

\bibliographystyle{aa} 
\bibliography{../../biblio2} 

%


\end{document}